\documentclass[amsmath,amssymb,twocolumn,pre,floatfix,superscriptaddress]{revtex4-2}
\usepackage[utf8]{inputenc}
\usepackage{graphicx}
\usepackage{color}
\usepackage{xcolor}
\usepackage{array}
\usepackage{hyperref}

\begin{document}

\title{Sequences of dislocation reactions and helicity transformations in tubular crystals}

\author{Andrei Zakharov}
\affiliation{Department of Materials Science and Engineering, University of Pennsylvania, Philadelphia, PA 19104, USA}
\affiliation{Department of Physics, University of California, Merced, CA 95343, USA}
\author{Daniel A.~Beller}
\email{d.a.beller@jhu.edu}
\affiliation{Department of Physics and Astronomy, Johns Hopkins University, Baltimore, MD 21218, USA}

\begin{abstract}

Freestanding tubular crystals offer a general description of crystalline order on deformable surfaces with cylindrical topology, such as single-walled carbon nanotubes, microtubules, and recently reported colloidal assemblies. 
These systems exhibit a rich interplay between the crystal's helicity on its periodic surface, the deformable geometry of that surface, and the motions of topological defects within the crystal. Previously, in simulations of tubular crystals as elastic networks, we found that dislocations in nontrivial patterns can co-stabilize with kinks in the tube shape, producing mechanical multistability. Here, we extend that work with detailed Langevin dynamics simulations, in order to explore defect dynamics efficiently and without the constraints imposed by elastic network models. Along with the predicted multistability of dislocation glide, we find a variety of irreversible defect transformations, including vacancy formation, particle extrusions, and ``reactions'' that reorient dislocation pairs. Moreover, we report spontaneous sequences of several such defect transformations, which are unique to tubular crystals. We demonstrate a simple method for controlling these sequences through a time-varying external force. 

\end{abstract}

\maketitle

\section{Introduction}

Freestanding tubular crystals represent a distinctive form of crystal structure found in both natural and technological settings. Characterized by their hollow, tube-like morphology, tubular crystals  consist of atoms or molecules arranged in a two-dimensional (2D) lattice structure closed along one direction into the topology of a cylinder. Unlike a 2D crystal attached to a cylindrical substrate \cite{lohr2010helical,mughal2012dense, wood2013self, tanjeem2021geometrical,liu2022curvature}, tubular crystals define their own tubular surface, allowing them to freely deform by distorting out of the tangent plane \cite{yakobson2001mechanical,BellerPRE16}.
Examples of freestanding tubular crystals include single-walled carbon nanotubes (SWCNTS) \cite{ajayan2001applications, baughman2002carbon}, capsids of filamentous viruses \cite{klug1999tobacco}, microtubules \cite{nogales2000structural}, and self-assemblies of DNA origami colloids \cite{hayakawa2022geometrically}. The emergence of such crystals has sparked significant interest due to their exceptional properties and their potential applications in nanoscience\cite{ajayan2001applications, baughman2002carbon}, functional materials \cite{wu2021application, wang2022advanced}  and biotechnology \cite{gu2022artificial}.

Similarly to other crystals, tubular crystals can contain lattice defects, such as grain boundaries and dislocations. These imperfections naturally arise due to various factors including crystal growth kinetics, lattice mismatch with supporting substrates, or external influences, and they play a significant role in determining a crystal's mechanical, optical, electrical and other properties \cite{dumitrica2006symmetry,zhu2016great}.  Controlling the number of defects and their dynamics is not only important for predicting crystal properties and tailoring them for applications, but it also provides pathways for synthesizing programmable structures \cite{zakharov2022shape}. In particular, the movement of dislocations alters the helicity of the tubular crystal, which in turn affects properties such as electrical conductivity \cite{yakobson2001mechanical} and growth kinetics \cite{artyukhov2014nanotubes, tanjeem2021geometrical}.  

In prior work \cite{zakharov2022shape}, we investigated the mechanics of dislocations in flexible, freestanding tubular crystals using a zero-temperature elastic network model. In that setting, dislocation test moves occurred through discrete, imposed bond flips in a network of harmonic springs, and the system evolved to minimize its total energy. This model predicted that dislocations could co-stabilize with kinks in the tube axis, introducing mechanical multistability in tube shape, such that certain imposed dislocation patterns allowed for programmable tube conformations and tunable mechanical response. 

However, the assumptions of the elastic network model exclude some phenomena likely to be of interest in colloidal tubular crystals. Because the model probes dislocation motions  by imposed bond flips, one defect at a time, it is challenging to investigate other types of possible particle rearrangements  such as vacancy formation, extrusion of particles from the crystal surface, and the interaction of existing dislocations with spontaneously nucleated defect pairs in dislocation ``reactions``.   

Here, we use Langevin dynamics simulations to investigate the spontaneous evolution of tubular crystals with thermal fluctuations and without constraints on the particles' contact network. Our simulated spherical particles interact through a model anisotropic interaction potential that stabilizes sheet-like assemblies. Such particles with designed surface pattern can be precisely engineered, for example by colloidal fusion \cite{gong2017patchy}, and they have demonstrated promise in the directed self-assembly of 2D and 3D structures using these particles as building blocks \cite{zhang2004self,van2006symmetry,duguet2016patchy,ravaine2017synthesis}. We initialize our systems with various tubular crystal configurations, including both chiral and achiral states. 

Although the elastic network model's predictions are often corroborated here, we additionally observe a wealth of more complex behaviors when dislocations pass near each other, including vacancy formation and particle extrusion. Most interestingly, we report novel dislocation reaction sequences that are unique to the tubular crystal setting, arising from the fact that distances between gliding dislocations change non-monotonically due to the surface's periodicity. We show that the interplay between dislocations, the helical symmetry of tubular crystals, and  out-of-tangent-plane deformations leads to a rich variety of transformations in crystal structure, including metastable dislocation patterns, dislocation reactions, and helicity transformations. 

\section{Methods: Particles and tubular assemblies}

\begin{figure}[t]
	\centering
	        \includegraphics[width=0.48\textwidth]{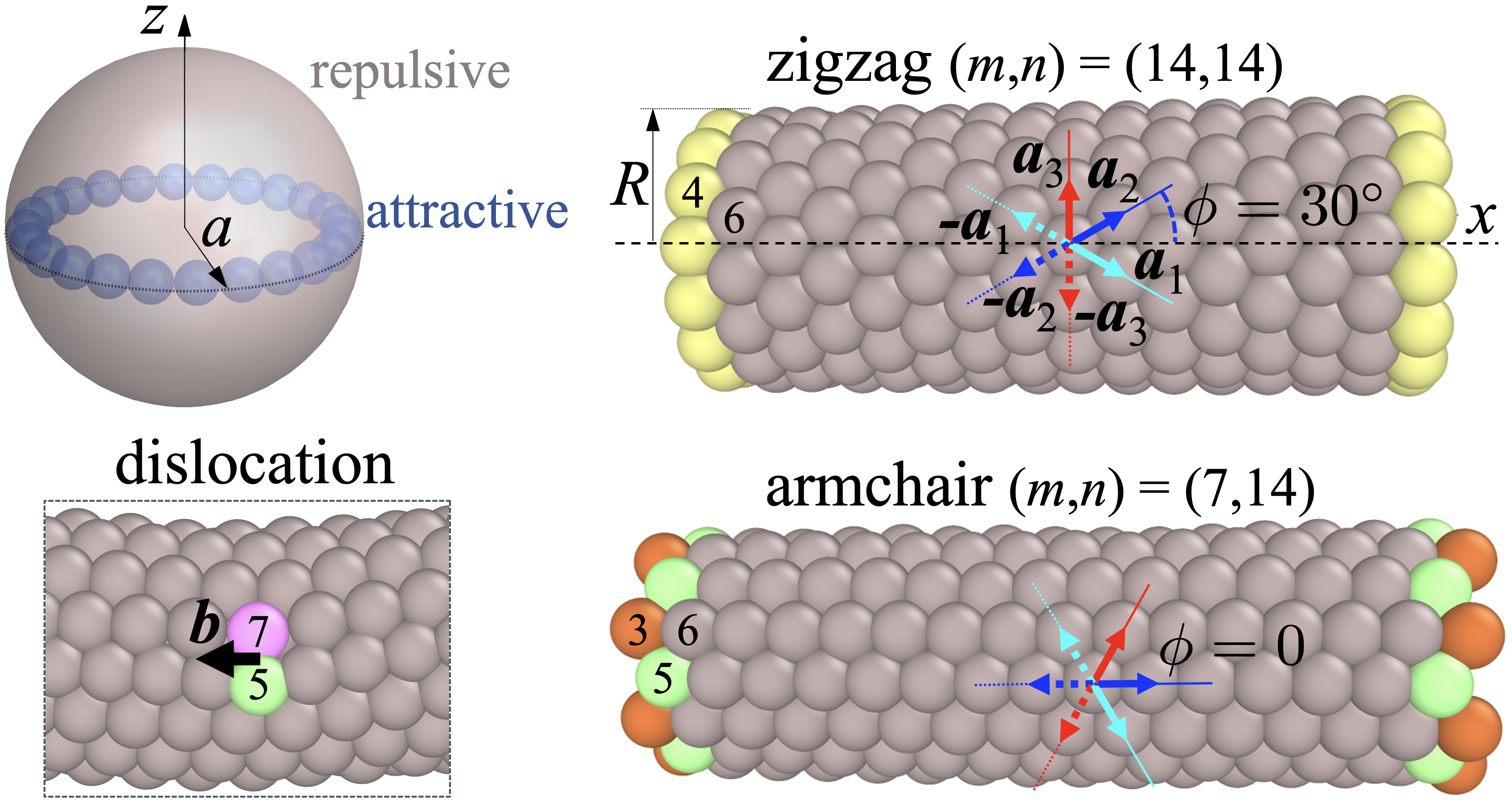}
	\caption{ Construction of freestanding tubular crystals. Tubular crystals (examples shown at right) are made up of patchy particles (upper left panel), each composed of a primary, sterically repulsive sphere and several smaller, attractive spheres. In the tube's local tangent plane, the particles form a close-packed hexagonal lattice, with primitive lattice directions $\pm \mathbf{a_1}$, $\pm \mathbf{a}_2$, $\pm \mathbf{a}_3$. The crystal's helicity is characterized by the angle $\phi$ between the tube axis $x$ and the steepest left-handed helical lattice direction or \textit{parastichy}, labeled $\mathbf{a}_2$ (upper right panel). Example tubular crystals shown are the achiral types: ``zigzag'', with $\phi=30^\circ$, and ``armchair'', with $\phi=0$. The number of distinct parstichies along the $\pm \mathbf{a}_1$ (the steepest right-handed helices) and along $\pm \mathbf{a}_2$ give the parastichy numbers $(m,n)$.  Particles are colored (and in some instances labeled) by coordination number, with gray identifying the 6-coordinated particles of a pristine hexagonal packing. A pair of 5- and 7-coordinated particles forms a dislocation (bottom left panel), and their separation vector is perpendicular in the local tangent plane to the dislocation's Burgers vector $\mathbf{b}$.}
	\label{Fig:Particle}
\end{figure}

 We construct tubular crystals from identical particles of spherical shape but with anisotropic interactions favorable for crystalline membrane assemblies. Neighboring particles interact as hard spheres except along a narrow equatorial band, where a mutual short-range attraction causes particles to stick together. To implement this patchy particle interaction, we model each particle as a rigid body composed of one purely repulsive {spherical ``atom'' of effective radius $a$}, together with a collection of attractive point-masses arranged along a circle at slightly smaller radius $a-\delta$, just under the larger sphere's surface (Fig.~\ref{Fig:Particle}). Repulsion between the larger spheres separated by center-to-center distance $r$ is determined by the Weeks-Chandler-Andersen (WCA) potential \cite{weeks1971role}, a truncated and shifted Lennard-Jones (LJ) potential \cite{Lennard_Jones_1931} given by
	\begin{align}
		E_r(r) &= \begin{cases}4\varepsilon_r \left[\left(\frac{\sigma}{r}\right)^{12}-\left(\frac{\sigma}{r}\right)^{6}\right]+\varepsilon_r,& \text{if } r\leq 2^{1/6}\sigma\\
    0,              & \text{otherwise}
  \end{cases}
		\label{Eq:LJ}
	\end{align}
where $\varepsilon_r$ is the interaction strength, and $\sigma=2a/2^{1/6}$ sets the cutoff distance to $r=2a$, equal to the minimum of the LJ potential, so that $E_r(r)$ is purely repulsive. 

The pair interactions between attractive atoms {on different particles} are defined by the soft potential 
	\begin{align}
		E_a(r) &= \begin{cases} 
            -\varepsilon_a [1+\cos(\pi r/r_a)],& \text{if $r\leq r_a$} \\
            0, & \text{otherwise}
            \end{cases}
		\label{Eq:softpot}
	\end{align}
with strength $\varepsilon_a$ and range $r_a$. For simplicity, we assume the attractive atoms are located along the equator of the repulsive atom, but in general they can have a more sophisticated arrangement \cite{zhang2004self}, such as a sinusoidal path. There is no interaction between repulsive and attractive atoms. 

We explore Langevin dynamics with these pair interactions by performing molecular dynamics (MD) simulations in LAMMPS \cite{plimpton1995fast}. We use scaled Lennard-Jones units: particles have unit mass ($m=1$), damping ($\gamma=1$), and steric repulsion strength $\varepsilon_r = 1$. The temperature is held fixed at $k_B T=0.1$ using a Nos\'e-Hoover thermostat. Because each particle is treated as a rigid body, the translation and rotation of the entire entity (large sphere and small spheres together) is calculated according to the net force and torque acting on its constituent atoms.

The anisotropy of patchy particles allows for their assembly into sheets \cite{zhang2004self}, and then into higher-energy tubular structures when a sheet is intentionally rolled up in one direction. In this work we do not consider the self-assembly process; instead, we construct initial conditions consisting of single-layer tubular crystals of given geometry, helicity, and preexisting defects in the lattice. In particular, we take as initial particle center positions the node locations of energy-minimized configurations obtained in our elastic network simulations \cite{zakharov2022shape}. Each particle is oriented so that its equatorial attractive band lies in the tube's local tangent plane. The particle separation $a$ is close to the particle diameter $2R$, so the particles in contact and stick together laterally. The imposed initial configuration is not necessarily a mechanical equilibrium state because dislocations can be in unstable positions. When the Langevin dynamics begin, the particle positions relax and, in many cases, we observe deformations in tube shape accompanied by motion of dislocations.

The tubular arrangement of particles can be conveniently described using the parastichy numbers, a pair of integers $(m,n)$ defining the number of distinct helices of particles in the steepest right-handed and steepest left-handed directions. In the family of helices there are three principal directions along unit vectors $\pm \mathbf{a}_i$, $i={1,2,3}$. {To determine helicity of a tubular assembly, we use the helicity angle $\phi$, which is }the angle between the steepest left-handed helix along the $\mathbf{a}_2$ direction and the tube axis (Fig.~\ref{Fig:Particle} upper right panel). Each elementary dislocation in the lattice is a distortion in the uniform particle tessellation, {composed} of a positive disclination at a five-coordinated particle and a negative disclination at a seven-coordinated particle. It can be characterized by a Burgers vector $\mathbf{b}$ of length $a$  orthogonal to the line connecting the five–seven disclination pair (Fig.~\ref{Fig:Particle} bottom left panel). 

We focus on tubes with an armchair and near-armchair crystal tessellation (one lattice direction approximately parallel to the tube axis) that are most common in the chirality family of carbon nanotubes (CNTs) \cite{artyukhov2014nanotubes}. Armchair lattices demonstrate large deformations  and dislocation pair nucleation when subjected to applied torsion \cite{yakobson1996nanomechanics}. In contrast,  zigzag tubes (the other achiral type) have one lattice direction oriented circumferentially and can more easily accommodate the shear stress caused by torsion via circumferential slip.

Chiral tubes of different  helicity with $0 < \phi < 30^\circ$ represent intermediate states between the two achiral cases, with a discrete transition step {$\Delta \phi = 30^\circ/m$}. While an exhaustive exploration of possible $\phi$-values is beyond the scope of this work, we focus here on certain small-$\phi$ tubular crystals to demonstrate the principle of controlled tubular crystal helicity transitions through dislocation interactions and reactions.

The constructed tubular {assemblies} are free-standing, and thus the tube local radius can change to reduce the energy associated with interparticle attraction and repulsion. For our patchy particles with equatorial attractive bands, the minimum energy configuration is a flat monolayer of particles with regular hexagonal closed packing. In tubular assemblies, every configuration has non-vanishing potential energy due to the curvature of the monolayer. The preference for a flat sheet morphology, encoded in the arrangement of attractive spheres on each particle's equator, introduces a bending energy, which was  imposed as a separate term in our previous elastic network model \cite{BellerPRE16, zakharov2022shape}. The bending energy depends on the tube radius, which is defined by the tessellation $(m,n)$ and the strength of particle interaction potentials. In our simulations we choose attractive potentials strong enough ($ \varepsilon_a=0.1$) to prevent disintegration of the structure due to the thermal fluctuations. On the other hand, we also avoid making the potentials so strong that the tube becomes faceted, with appearance of planar regions joined at creases. At very large effective bending stiffness, this faceting helps to reduce the overall bending energy by localizing it at the creases \cite{zandi2004origin, vernizzi2007faceting,garcia2021faceting}.  The effective bending energy associated with curvature of a tubular structure of length $L$ and radius $R$ can be approximated as $F_{bend}\approx \pi \kappa L/R$, where $\kappa$ is the effective bending stiffness. Thus, increasing the tube radius will lead to a lower energy configuration.

\section{Results}

\subsection{Dislocations spontaneously react in a free-standing tubular crystal}

\begin{figure}[t]
	\centering
			\includegraphics[width=0.5\textwidth]{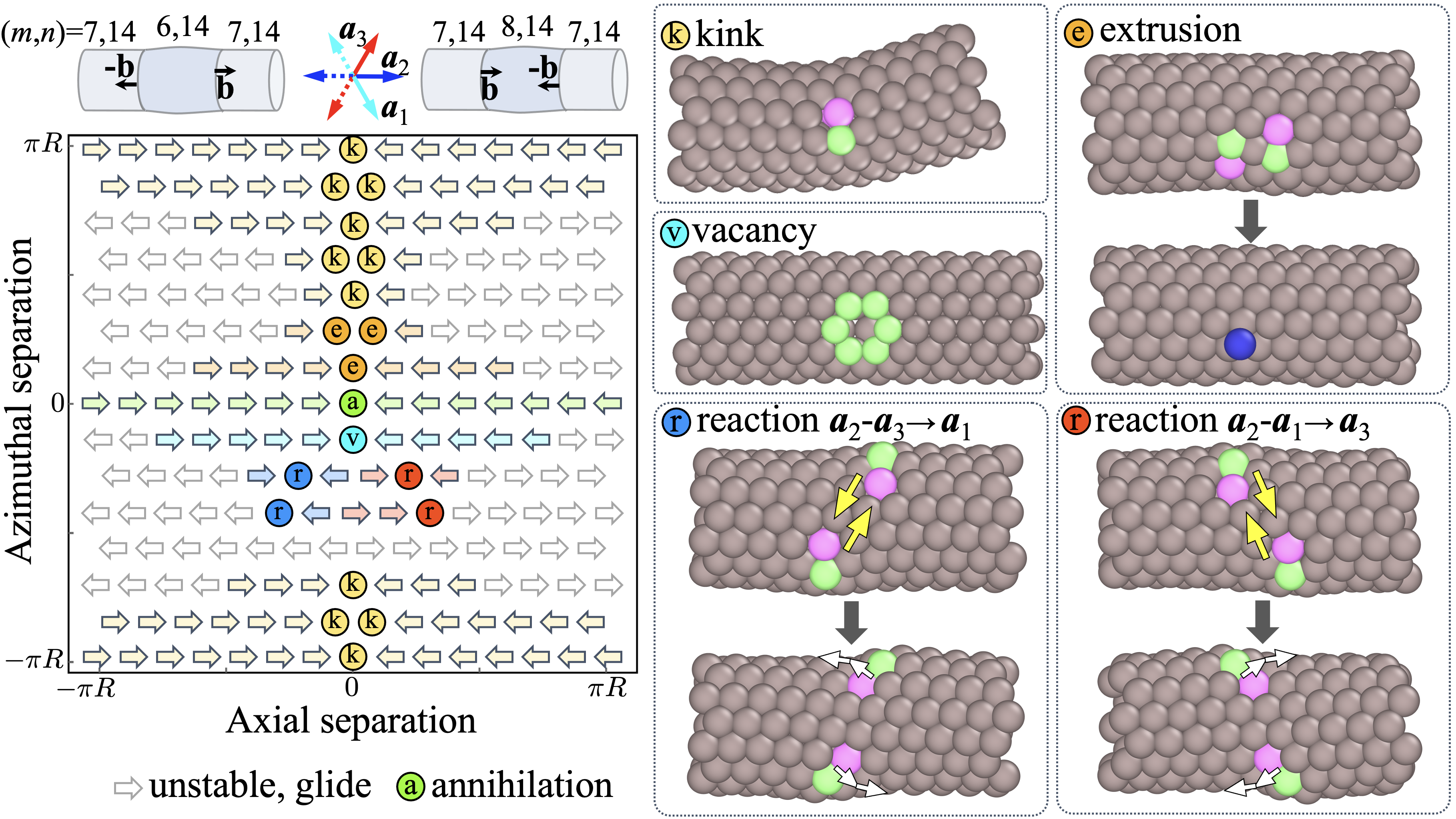}
	\caption{A stability map for two interacting dislocations oriented along the tube axis ($\pm \mathbf{a}_2$ parastichies) in an armchair tube ($(m,n) = (7,14)$, $\phi= 0$) at different values of azimuthal (climb) separation. The region of the tube between the dislocations has a different tessellation, $(m,n)=$ (6,14) or (8,14), due to the helicity transition at the dislocations.  Unstable states are indicated  by wide arrows, which are colored if the system is attracted to a metastable state labeled with the corresponding color. If the axial separation is larger than $\pi R$ (outside the plotted region), dislocations interact weakly and are always repulsive, simply gliding toward the tube ends. Labels for metastable state types correspond to representative metastable tube configurations depicted at right. } 
	\label{Fig:2disl_0}
\end{figure}

First, we explore stability of a free-standing tubular crystal with armchair $(m,n)=(7,14)$ arrangement of particles and containing two oppositely oriented,v isolated dislocations with Burgers vectors $\pm \mathbf{b}$ (Fig.~\ref{Fig:2disl_0}). This configuration, while containing a net-zero Burgers vector ``charge'',  requires a phyllotactic transition in which $(m,n)$ differs in the region between the dislocations compared to outside them, by an amount depending on the orientation of $\mathbf{b}$. 

In Fig.~\ref{Fig:2disl_0} we plot a stability map describing the evolution of two dislocations with Burgers vectors along (or opposite to) the axial direction, depending on the axial component $g\cdot a$  and azimuthal  component $c\cdot a$ of separation between the  defects, where $g$ and $c$ are integer numbers of glide and climb steps, respectively. {When dislocations move, they do so exclusively in glide motions along their Burgers vectors, as this is energetically cheaper than climb motion along other directions and requires no change in total number of particles.} As the Burgers vectors lie along the tube axis in this example, the defect motions recorded in Fig.~\ref{Fig:2disl_0} are only along the axial direction; the azimuthal separation remains fixed. For larger azimuthal separations the dislocations glide to decrease their axial separation to zero.

Even though the middle region of the tube has a larger radius with lower bending energy, the dislocations are attracted to a state with minimum axial separation at most possible azimuthal separations, which produces a kink in the tube axis as observed in the elastic network model \cite{zakharov2022shape}. This interaction is observed for axial separations close to $\pi R$ when the azimuthal separation is around $\pi R$, but the interaction range decreases as azimuthal separation decreases. 

The outcome changes significantly if the  azimuthal (climb) separation is small, $|c| \leq 3$. We can divide this regime into two parts, one with $c>0$ where the 5-coordinated particles are closer to the other dislocation (``fives inside''), and one with $c<0$ where the 7-coordinated particles are closer (``sevens inside'').    For the fives inside case, the two dislocations attract to zero axial separation and then disappear through an \emph{extrusion} event, in which one or more particles are expelled from the surface of the tubular crystal and forced to sit on top of it, at larger radius. For $c=1$, a single particle is extruded, becoming a 3-coordinated particle (colored in blue in the ``extrusion'' panel of Fig.~\ref{Fig:2disl_0})  adjacent to three 7-coordinated particles under it (colored in gray to show pristine arrangement in the tube tangent plane). The total number of bonds is conserved in this extrusion. Dislocations separated by $c=2$ climb steps cause an extrusion of two particles at small axial separation.   

An opposite azimuthal separation, with $c=-1$ and ``sevens inside'', leads instead to the formation of a vacancy. The total number of bonds decreases, leaving six five-coordinated particles that become a nucleation site for plastic deformations when the tube is under applied force. For example, applied torsion can destabilize this metastable vacancy, replacing it with two dislocations oriented along a lattice direction that depends on the sign of the torsion: Simulations show that unlike a pristine tube under torsion, in which unbinding dislocations have Burgers vectors $\pm \mathbf a_2$, the preferable path for nucleation in a tube with a vacancy is along $\pm \mathbf a_1$ when the torsion is applied in the counter-clockwise direction, and along $\pm \mathbf a_3$ when torsion is in the clockwise direction [Supplemental Material Video 1-2].

While the vacancy formation at $c=-1$ is intuitively opposite to the extrusion at $c=+1$, we find an altogether new behavior for $c=-2$ or $-3$: When the gliding dislocations reach points that sit on the same $\pm \mathbf{a}_3$ parastichy, their Burgers vectors appear to suddenly change from $\pm \mathbf{a}_2$ to $\pm \mathbf{a}_1$. This is a dislocation \emph{reaction} event \cite{gerbode2008restricted,kubin2013dislocations,irvine2013dislocation}. 
As we explore in detail below, we can understand this process as resulting from the nucleation of a new dislocation pair with Burgers vectors $\pm \mathbf a_3$. The new dislocations glide apart along the $\pm \mathbf{a}_3$ parastichy until each new dislocation reacts with one of the original dislocations, leaving behind two dislocations with resultant Burgers vectors $\pm \mathbf{a}_2 + (\pm \mathbf{a}_3) = \pm \mathbf{a}_1$. After the reaction, the $\pm \mathbf{a}_1$ dislocations glide only a few steps before coming to rest in a metastable state. The mirror-image process is also observed, with dislocations of initial Burgers vector  $\pm \mathbf a_2$  and separation  along a $\pm \mathbf a_1$ parastichy reacting to become dislocations with Burgers vectors $\pm \mathbf{a_3}$.

In this tube with $(m,n)=(7,14)$, there is only a single azimuthal separation, at $c=4$ climb steps, at which the dislocations are always repulsive and do not react. At zero azimuthal separation and axial separation less than $\pi R$, dislocations are attractive and annihilate, forming a pristine tube. 

\begin{figure}[t]
	\centering
			\includegraphics[width=0.485\textwidth]{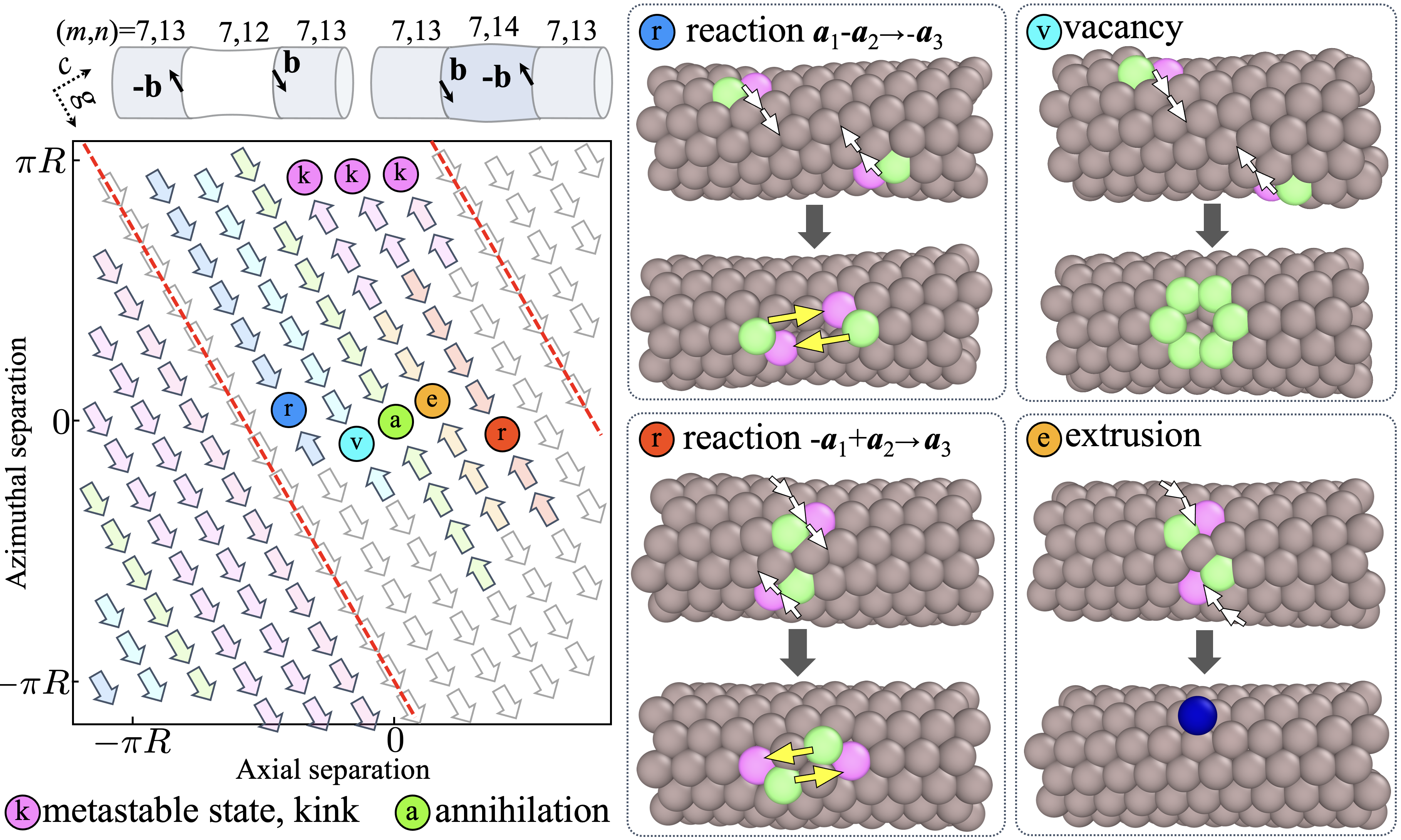}
	\caption{A stability map for two interacting dislocations spontaneously gliding along helical trajectories ($\pm \mathbf{a}_1$ parastichies) in a near-armchair tube with $(m,n)=$ (7,13), $\phi\approx 2.5^{\circ}$, at various values of climb separation $c$. If the dislocations are initially at negative axial separation, then the central region of the tube has $(m,n)=(7,12)$ and decreased radius, and the dislocations are attracted to one or more metastable states at small axial separation. The exception to this rule is  a single path at $c=-3$ (a line with empty arrows {highlighted with a dashed red line}). }
	\label{Fig:2disl_pi3}
\end{figure}

More generally, the glide paths of dislocations in tubular crystals are helices rather than straight lines, a distinction that we find has important consequences for defect interactions. In Fig.~\ref{Fig:2disl_pi3} we show a stability map for a near-armchair tubular crystal, $(m,n) = (17,13)$, with two dislocations whose Burgers vectors  ($\mathbf{b}_{1,2} = \pm\mathbf{a}_1$) make the glide paths helical. At large positive or negative axial separation $\Delta x$, the dominant trend is simply for $\Delta x$ to steadily increase. Unlike the previous example with dislocations along the tube axis ($\pm\mathbf{a}_2$), here the helicity transition $(\Delta m, \Delta n)$  induced by these defects' relative glide motion causes regions of larger radius to grow, at the expense of regions of smaller radius, as $\Delta x$ increases. 

However, once dislocations reach smaller axial separation, they usually interact in a more complex way that stabilizes their relative positions. We find two groups of these metastable states at small $\Delta x$. In the first group, dislocation pairs come to rest at maximum azimuthal separation $\Delta y$ (i.e.\ opposite sides of the tube), giving the tube a kinked conformation.  The other group,  at small azimuthal separation, results in more dramatic changes to the defects, strongly depending  on the climb separation $c$ between the two glide paths. For $c=0$, the dislocations annihilate, just as in the previously examined case of $\mathbf{b}$ parallel to the tube axis. For $|c| = 1$, the final state has no dislocations but contains either a vacancy ($c=-1$, sevens inside) or a single extruded particle $(c=+1$, fives inside).  

Examining the $|c|=2$ paths, we find that dislocation reactions may occur for both sevens-inside and fives-inside configurations, in contrast to the previous example with Burgers vectors parallel to the tube axis. The reactions in this case replace the $\pm \mathbf{a}_1$ Burgers vectors with new ones along $\pm \mathbf{a}_3$. For $c=+2$, there are two different metastable states along the same glide path: the ``reacted'' state with two nearby $\pm \mathbf{a}_3$ dislocations in an approximately straight tube, and a ``kinked'' state with diametrically opposed $\pm\mathbf{a}_1$ dislocations. 

For this particular choice of initial parastichy numbers  $(m,n) = (7,13)$, there is only one climb separation $c=-3$ {(highlighted with a dashed red line along the helical path in the stability map in Fig.~\ref{Fig:2disl_pi3})} that allows two dislocations to ``miss'' each other, gliding past $\Delta x=0 $ to $\Delta x \rightarrow + \infty $ without any metastable state. We expect that the  number of such free paths will be greater for tubes with larger $(m,n)$.

\subsection{Inducing dislocation movements and reactions through applied torsion}

\begin{figure}[t]
	\centering
		\includegraphics[width=0.485\textwidth]{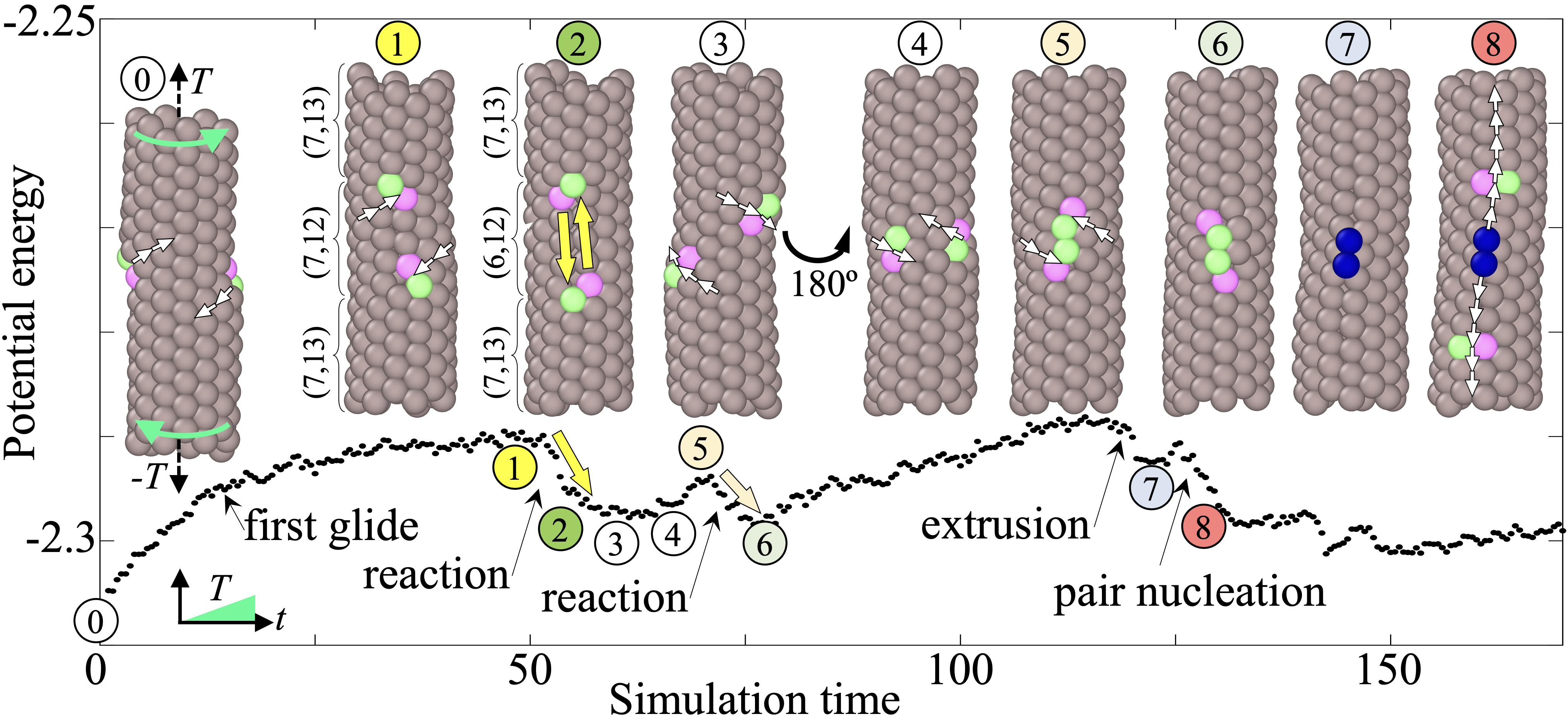}
	\caption{Evolution of the potential energy  during a sequence of spontaneous dislocation glide, reaction, extrusion, and pair-unbinding events in a near-armchair tube under applied torsional force $T$. The initial state 0 is metastable with two dislocations at angle $\theta\approx-\pi/3$ to the tube axis. Applied torsion linearly increases with time. The final state has two extruded particles (blue),  an altered helicity $(m,n)=(6,13)$ throughout the tube, and no dislocations once those in state 8 glide apart to the ends of the tube. The view of the tube is rotated $180^\circ$ between states 3 and 4, and is otherwise held fixed. }
 \label{Fig:torsion_2disl}
\end{figure}

To better understand how helicity transitions in tubular crystals may be controlled, we next examine the effects of externally imposed torsion applied at the ends of the tube. Just as an applied shear stress provides control over  an isolated dislocation's motion in a planar crystal \cite{irvine2013dislocation}, here torsion imposed at the tube ends produces shear stress in the tube's tangent plane and thus a Peach-Koehler force on the dislocations. Because we are particularly interested in switching between mechanically metastable states, we study in detail  the multistable glide path identified above in the $(m,n)=(7,13)$ tube, with  $c=+2$ climb separation. We find that we can switch between the reacted and kinked metastable states by means of this applied torsion. Moreover, the transition is reversible through reversal of the torsion direction. While the kinked state has higher energy than the reacted state, we can impose a torsion sufficiently strong that the ``reacted'' dislocations undergo a reverse reaction from their $\pm \mathbf{a}_3$ Burgers vectors back to their original  $\pm\mathbf{a}_1$ orientations and then glide to the metastable kinked state.

To quantitatively analyze this mechanical reconfigurability, we plot in Fig.~\ref{Fig:torsion_2disl} the change in the tube's total potential energy as the applied torsion is increased monotonically with time. We take the $c=+2$ metastable kink state of Fig.~\ref{Fig:2disl_pi3} as the initial configuration and apply a constant torsion $T=250$ (illustrated in the 	state 0 panel of Fig.~\ref{Fig:torsion_2disl}) at the ends of tube. As simulation time progresses, we find  not only an escape from the metastable kink state into the reacted state, but also a series of subsequent events including extrusion and other reactions. Representative states from this sequence of transitions are shown as inset panels to Fig.~\ref{Fig:torsion_2disl}. 

At first, applied torsion leads to an increasing potential energy and destroys the metastable state, inducing glide (at $t=15$), which reduces the slope in the energy increase. As  the dislocations glide apart along helical trajectories, they increase their axial separation but eventually bring their azimuthal separation to near zero (state 1 at $t=50$, Supplemental Material Video 3). At that time, with the dislocations located on the same $\pm \mathbf{a}_2$ parastichy, a dislocation reaction event occurs rapidly. The reaction changes the two Burgers vectors from $\pm\mathbf{a}_1$ to $\pm\mathbf{a}_3$ and thus changes the parastichy numbers describing the middle region of the tube from $(7,12)$ to $(6,12)$ (state 2 at $t=55$). The potential energy drops sharply with this reaction event. Subsequently, with their new helical $\pm \mathbf{a}_3$ glide paths, the dislocations continue to glide with the azimuthal components favored by the applied torsion while also decreasing their axial separation, which allows a reduction in the size of the tube's narrower central region and thus a reduction in bending energy. Eventually, the dislocations approach each other again on the opposite side of the tube (state 4 at $t=70$), where they react a second time and obtain Burgers vectors $\pm\mathbf{a}_1$ (state 6 at $t=80$). In this state, the two positive disclinations are in contact, creating a large compressive force that obstructs further motions of the dislocations for some time. Continuing applied torsion causes a gradual increase in potential energy and, eventually, an extrusion of two particles takes place (state 7 at $t=120$). Finally, the extruded particles become a nucleation point for a pair of dislocations along the $\pm\mathbf{a}_2$ directions (state 8 at $t=130$). 

This example demonstrates that dislocation reactions decrease the elastic energy in a tubular crystal through reorientation of existing dislocations when torsion is applied.  Although the hexagonal crystal structure imposes limitations on the available glide motion and reaction directions, specifically along the three lattice directions $\pm\mathbf{a}_{1,2,3}$, the tube's periodicity enables a sequence of reactions. In planar crystals, the spatial separation between a pair of dislocations varies monotonically with their glide motion. In contrast, the tube's periodicity facilitates multiple reaction events: After the initial reaction, dislocations glide apart along helical paths but eventually become closer again on the opposite side of the tube, leading to further reactions.

\subsection{An analytical prediction for dislocation motion and reactions on a cylinder under shear stress}

\begin{figure}[t]
	\centering			\includegraphics[width=0.485\textwidth]{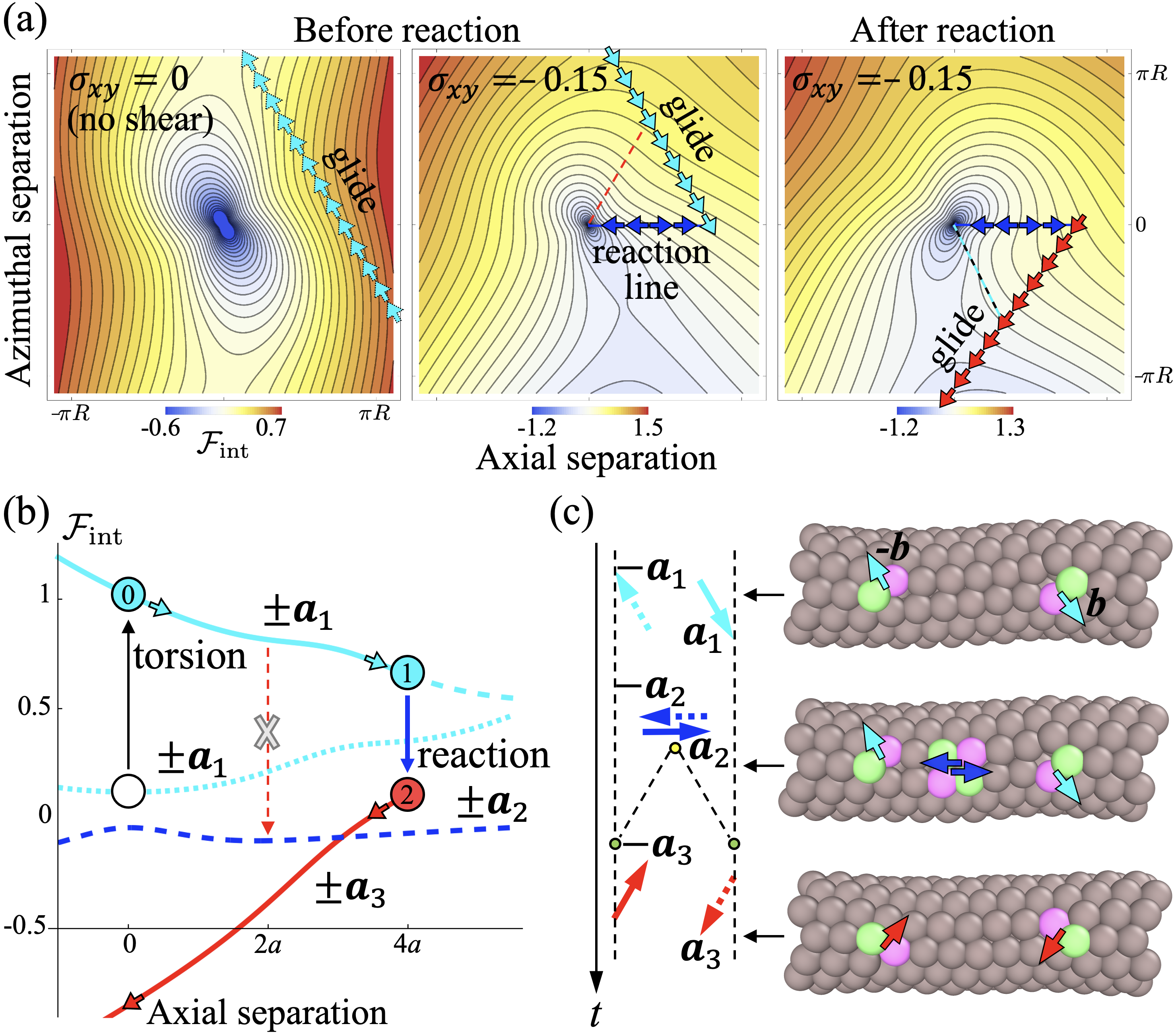}
	\caption{Analytical approximations for interaction energies of reacting dislocations on a tubular crystal. (a) Calculated interaction energy, $\mathcal{F}_\mathrm{int}$, and the relative glide direction of a pair of dislocations. The climb separation is constant $c=4a\sqrt{3}/2$ but the orientation of the Burgers vectors changes from $\pm \mathbf{a}_1$ (before reaction) to $\pm \mathbf{a}_3$ (after reaction). Without applied shear stress ($\sigma_{xy}=0$), dislocations are attracted to a stable state at zero axial separation. That state is destabilized by a shear stress $\sigma_{xy}=-0.15$, which causes the dislocations to glide apart and eventually react at zero azimuthal separation. Subsequently, the rotated defects glide in a different direction. (b) Interaction energy along the glide paths depicted in (a). The metastable state (white circle) in the zero-torsion path (cyan dotted line) becomes unstable when torsion is applied, and dislocations glide (cyan solid line) from state 0 to state 1, where they react into a new state 2 and then glide along another glide path (red solid line). Possible but energetically unfavorable glide paths and reactions are shown by dashed lines and arrow, respectively. (c) A schematic diagram and representative states for the reaction event. In a tube under torsion, an extra pair of dislocations spontaneously nucleates and quickly unbinds between two existing dislocations, gliding toward and eventually reacting with them. After the reaction, the tube contains one pair of dislocations but with altered Burgers vector orientations,  and the helicity of the region between the dislocations is altered. 
    }
	\label{Fig:shear_2disl}
\end{figure}

To better understand why we observe some possible reactions and not others in MD simulations, we examine the interaction energy between two dislocations using an analytical approximation for dislocation pair energy on a cylinder calculated in \cite{Amir13}. The calculation omits the Peierls barrier to glide steps and assumes that the crystal is wrapped around a rigid cylindrical surface, in contrast to the freestanding tubular crystals in our study. Additionally, the bending energy favoring larger tube radius plays no role in the rigid-cylinder calculation. Nonetheless the comparison is qualitatively illuminating when we examine one-dimensional cuts, representing parastichies, through the predicted two-dimensional interaction energy landscapes. 

In Fig.~\ref{Fig:shear_2disl}a we present interaction energy landscapes for two  dislocations with Burgers vectors $\pm \mathbf{a}_1$ at constant angles to the tube axis $\theta=-\pi/3$, $2\pi/3$, before and after application of an external shear stress $\sigma_{xy}$. Without shear stress (Fig.~\ref{Fig:shear_2disl}a left panel), dislocations at any position along the glide path will glide (arrows) to a stable state with minimal axial separation to minimize the bending energy. Although the analytical theory \cite{Amir13,BellerPRE16} accounts only for the in-plane stresses, it also correctly predicts the direction of dislocation glide that leads to the family of states at small axial separation.

Applied shear stress changes the energy landscape, destabilizing the states at small axial separation, and predicting glide motion that increases the axial separation (cyan arrows in Fig.~\ref{Fig:shear_2disl}a middle panel). Because our dislocations have initial Burgers vectors $\pm \mathbf{a}_1$, reactions are possible when glide along their $\pm \mathbf{a}_1$ parastichies brings them to the same $\pm \mathbf{a}_2$ or $\pm \mathbf{a}_3$ parastichy, which we mark with dashed red line and blue arrows. Even though the common $\pm \mathbf{a}_3$ parastichy is reached first, we do not observe a reaction {along this path} in our MD simulation. This observation is consistent with a  prediction given in Ref.~\cite{BellerPRE16} for the Burgers vectors of dislocation pairs expected to spontaneously unbind in the presence of external torsion. In particular, for a tube with helicity angle $\phi=0$ as in Fig.~\ref{Fig:shear_2disl}, the $\pm \mathbf{a}_2$ direction (blue arrows) is more favorable for unbinding than the $\pm \mathbf{a}_3$ direction (cyan dashed line). As a result, the original dislocations glide along $\pm \mathbf{a}_1$ until reaching a common $\pm \mathbf{a}_2$ parastichy with minimal azimuthal separation, and then react along the $\mathbf{a}_2$ direction. This reaction leads to reorientation of the Burger vectors by $\pi/3$ to a new pair of directions $\pm\mathbf{a}_3$. The calculation of Ref.~\cite{Amir13} then predicts a different interaction energy landscape, which we show along with the new glide path (marked by 
red arrows) in the right panel of Fig.~\ref{Fig:shear_2disl}a.

The energetic reasons for the observed dislocation glide and reaction behaviors become more apparent when we take one-dimensional cuts of the energy along the relevant parastichy paths. In Fig.~\ref{Fig:shear_2disl}b we plot the analytically predicted interaction energy along the glide paths before (cyan dotted line) and after applying an external shear stress (cyan and red solid lines). Whereas an azimuthal separation of $-\pi R$ is metastable without applied torsion (open circle), the applied torsion creates a slope in the interaction energy that produces a force causing dislocations to glide from state 0 to state 1 (cyan solid line), corresponding to the glide motion along helical trajectories observed in our MD simulations (Fig.~\ref{Fig:torsion_2disl}). Along the path from state 0 to state 1, there is a configuration with possibility for dislocations to react and change their orientation to $\theta=0$ (red dashed line). Although this possible reaction would substantially decrease the predicted interaction energy, the energy barrier associated with unbinding along the path is large according to the prediction in Ref.~\cite{BellerPRE16}; thus the reaction does not happen, and dislocations instead proceed to state 1. The reaction event at state 1 leads to reorientation of the Burgers vectors from $\theta=-\pi/3$ to $\theta=-2\pi/3+\Delta$, where $\Delta$ is a small angle due to the helicity transition. This reorientation of Burgers vectors is also associated with a decrease in the value of interaction energy (red solid line in Fig.~\ref{Fig:shear_2disl}b) and a more negative slope of the energy with respect to glide (which now decreases rather than increases the axial separation). For comparison, we also plot the higher predicted interaction energy for the hypothetical situation in which the dislocations continue along $\pm \mathbf{a}_1$ without reacting (dashed cyan line in Fig.~\ref{Fig:shear_2disl}b). 

\subsection{Mechanism of dislocation reaction}

To examine dislocation reactions in tubular crystals in greater detail,
we increase the climb separation between the initial dislocations of Fig.~\ref{Fig:shear_2disl}a,b and apply external torsion to move the dislocations to a configuration where they undergo a reaction. This allows us to observe the mechanism of the reaction, and resulting changes in lattice helicity, over a larger region. In Fig.~\ref{Fig:shear_2disl}c we show simulated configurations and schematic representations of the dislocations before, during, and after the reaction. Due to applied torsion, the dislocations with initial Burgers vectors $\pm \mathbf{a}_1$ glide apart until  reaching a common $\pm \mathbf{a}_2$ parastichy. Then, we observe that a new dislocation pair with Burgers vectors $\pm \mathbf{a}_2$ nucleates midway along the common parastichy, with the net Burgers vector remaining zero, as it must.  While the original dislocations remain stationary, the newly nucleated dislocations unbind and glide toward the original defects until eventually combining with them. The reaction ends with a new configuration containing two dislocations with opposite Burgers vectors $\pm\mathbf{a}_3$ direction. This reoriented dislocation pair is not necessarily stable at its current separation, so the defects may next glide along their new $\pm \mathbf{a}_3$ helical glide paths to decrease the total energy.

We never observe dislocations with $|\mathbf{b}| > a$ in our simulations; all of the dislocations are elementary. A consequence of excluding larger Burgers vectors is that reactions by the mechanism we describe can never occur along the $\mathbf{b}$ direction, even if the dislocations sit on the same glide parastichy ($c=0$), because a nucleated dislocation pair could only glide along such a path if its Burgers vectors were parallel to those of the original dislocation pair; the reaction would then produce non-elementary dislocations. (Geometrically, a reaction could occur that annihilates all dislocations, but this seems energetically implausible compared to the original dislocations simply gliding to zero separation and annihilating there.)  

The final Burgers vectors are explicitly defined by the initial defects' orientations and the reaction-nucleated dislocations. For two oppositely oriented dislocations along $\pm\mathbf{a}_1$, a dislocation pair unbinding is possible along the other two primitive lattice directions $\pm\mathbf{a}_{2,3}$. The favorable direction for unbinding dislocations under torsional stress depends on the helicity angle $\phi$ \cite{BellerPRE16}. For dislocations initially along $\mathbf{a}_1$, reactions will transform $\mathbf{a}_1 + \mathbf{a}_2 \rightarrow \mathbf{a}_3$ or $\mathbf{a}_1 + \mathbf{a}_3 \rightarrow \mathbf{a}_2$ (omitting the signs), where the energy barrier associated with unbinding along $\mathbf{a}_3$ direction is larger than along $\mathbf{a}_2$, and thus is not favorable, as predicted in Ref.~\cite{BellerPRE16}.

\subsection{Controlling helicity through dislocation reactions and alternating torsion}

\begin{figure}[t]
	\centering
	\includegraphics[width=0.485\textwidth]{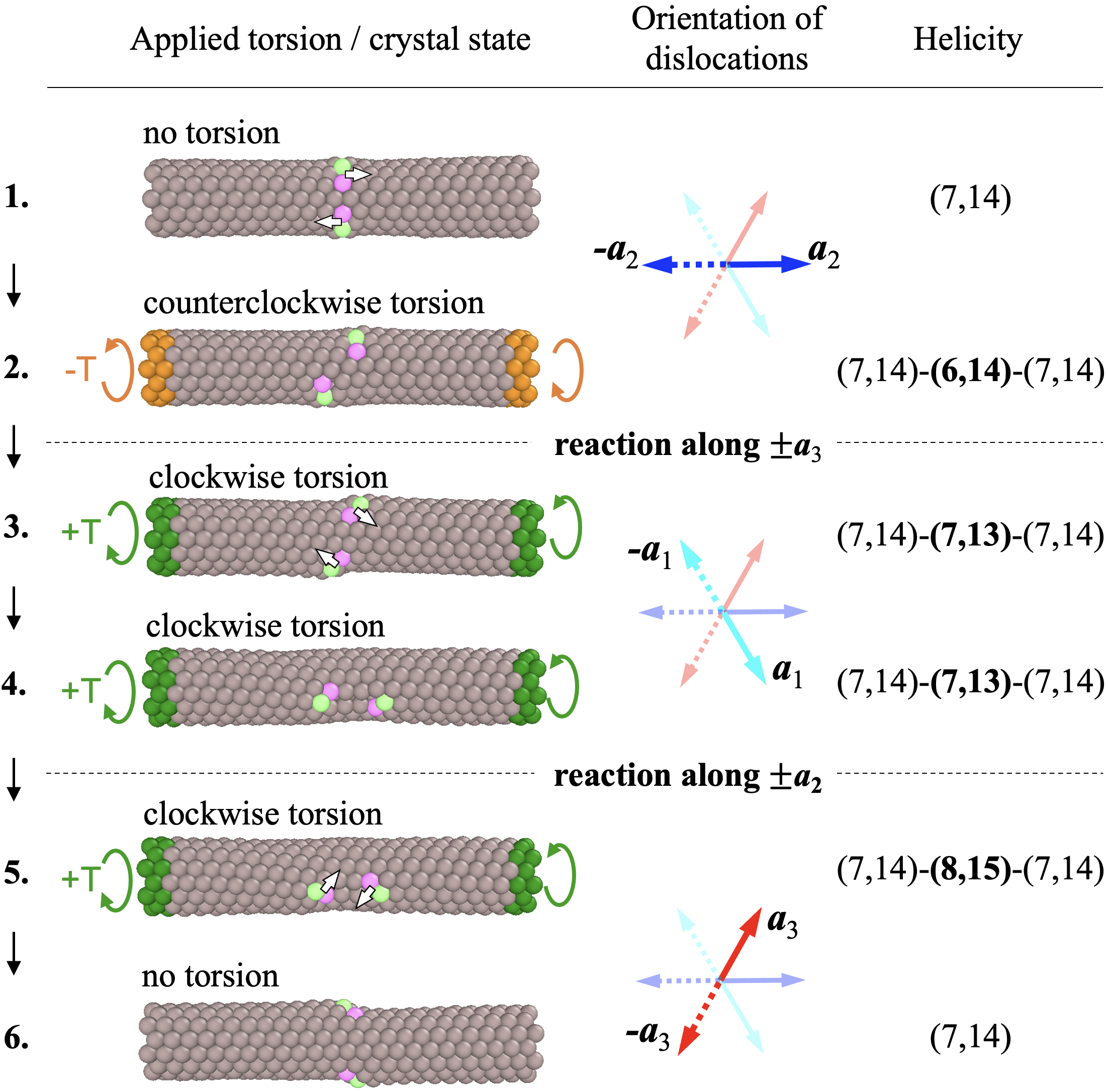}
	\caption{Transitions in dislocation orientation and crystal helicity through a sequence of reactions controlled by time-varying external torsion. Applying torsion in the clockwise direction to the initial configuration with two dislocations along $\pm \mathbf{a}_2$ parastichy (state 1) leads to a dislocation glide (state 2) and a reaction along $\pm \mathbf{a}_3$ direction. Switching the direction of applied torsion (state 3) moves repulsive dislocations closer to each other again (state 4), which is followed by a reaction along the $\pm \mathbf{a}_2$ parastichy and a change of Burgers vectors to the $\pm \mathbf{a}_3$ direction (state 5). Finally, dislocations glide to a metastable state at maximum azimuthal separation (state 6). Particles subject to applied torsion in counterclockwise or clockwise direction are colored in orange and green, respectively. }
	\label{Fig:Series}
\end{figure}

Having demonstrated that externally imposed torsion can move dislocations into configurations where they react, we further demonstrate here that time-dependent torsion of alternating sign can be used to change the helicity of the tubular crystal. Such helicity changes hold significant practical implications, particularly in the field of carbon nanotubes (CNTs) to control their electric conductivity.  Here we start with an achiral configuration of crystal with an armchair $(7,14)$ tube containing two oppositely oriented dislocations along $\pm \mathbf{a}_2$, initially separated by three lattice spacings in the azimuthal direction (Fig.~\ref{Fig:Series}). As shown in Fig.~\ref{Fig:2disl_0}, this state is unstable and dislocations spontaneously glide apart to a state where they react along a $\pm \mathbf{a}_1$ parastichy, changing their orientation to $\pm \mathbf{a}_3$. However, applying shear stress by means of torsion on the tube ends, we can prevent the reaction and instead force the defects to move in the opposite direction to a state (stage 2 in Fig.~\ref{Fig:Series}) which makes possible a reaction along a $\pm \mathbf{a}_3$ parastichy. If the applied torsion is removed immediately after this reaction, dislocations will remain in this orientation and move apart to a state with maximum azimuthal separation. By reversing the direction of torsion (stage 3), we then cause dislocations to glide toward a state where they can react along a $\pm \mathbf{a}_2$ parastichy (stage 4) and change their orientation once again, this time to the third possible direction $\pm \mathbf{a}_3$ (stage 5). Finally, upon removal of the applied torsion, the dislocations glide to maximize their azimuthal separation and reach a metastable state (stage 6, Supplemental Material Video 4). 

\section{Conclusion}

In order to elucidate the complex interplay between topological defects, lattice chirality, and surface geometry in freestanding tubular crystals, we have employed MD simulations to investigate the behavior of spherical patchy particles, organized into tubular structures with pre-existing defects. In agreement with predictions made using elastic network simulations \cite{zakharov2022shape}, our study finds the emergence of stable dislocation patterns, a behavior specific to tubular crystals, causing significant deformations in the macroscopic shape of the crystal. Beyond the scope of the previous elastic network approach, our MD simulations here also reveal novel sequences of dislocation reaction events that are distinctive to tubular crystals. Even when the existing defects are positioned several lattice spacings apart, they exhibit reactions ``at a distance'', through nucleation and glide-separation of an intervening dislocation pair, that lead to changes in their orientations. Consequently, this alteration in the orientation of dislocations induces transformations in the helicity of the lattice. In addition, the vacancy formation and particle extrusion events observed in our molecular dynamics simulations represent a class of irreversible rearrangements, not accessible to the elastic network model, and could provide target sites for hierarchical assembly bonds or nucleation sites for secondary crystalline layers.

Our investigation establishes externally applied torsion as a promising means to effectively regulate the dislocation motion. We show that time-varying manipulation of the direction and magnitude of these forces can be used to initiate a sequence of elementary reactions that result in a desired dislocation reorientation, defect pattern, and change in crystal helicity. This controllable design of dislocation dynamics holds the potential to engineer colloidal crystal assemblies with \textit{in situ} tunable mechanical and electro-optic properties. Despite the strong assumptions underlying the analytical theory of Ref.~\cite{Amir13}, we found that it generated useful predictions for dislocation glide both before and after a dislocation reaction. Our findings indicate a need for an extension to this theory that can systematically predict which geometrically possible dislocation reactions will be energetically favorable. 

Our results suggest that it would be fruitful in future work to examine dislocation reaction sequences on 2D crystals of other periodic topologies, such as spherical and toroidal, and on tubular crystals with helicites other than the (near-)armchair configurations that we have studied here. How these dislocation reaction sequences may influence the kinetics of tubular crystal self-assembly remains an open question. Slight variations to our patchy particle construction could offer future simulations a minimal model of  tubular crystals with spontaneous curvature, preferred helicity, and/or crystal symmetries other than hexagonal, modeling the rhombic lattices of tubulin that make up microtubules \cite{cheng2014self} or potential colloidal analogues \cite{li2005fabrication}. 

Altogether, our findings not only offer insights into the rich interplay between topological defects, lattice helicity, and surface geometry at play in freestanding tubular crystals, but also advance our understanding of dislocation behavior in flexible 2D crystals and the variety of ways in which they can interact. 

\section*{Acknowledgements}

AZ acknowledges computing time on the Multi-Environment Computer for Exploration and Discovery (MERCED) cluster at UC Merced, which was funded by National Science Foundation Grant No. ACI-1429783. Part of this research was also conducted using Pinnacles cluster (NSF MRI, 2019144) at the Cyberinfrastructure and Research Technologies (CIRT) at University of California, Merced.

\bibliography{MD.bib}

\clearpage

\section*{Supplemental Material Video}

\textbf{Video 1.} Applied torsion in the counter-clockwise direction to a tubular crystal with a preexisting vacancy defect leads to unbinding dislocations that travel along helical paths opposite to the direction of torsion.
\\

\textbf{Video 2.} Applied torsion in the clockwise direction to a tubular crystal with a preexisting vacancy defect leads to unbinding dislocations that travel along helical paths in the counter-clockwise direction.
\\

\textbf{Video 3.} A tubular crystal, with two initial dislocations gliding on helical parastichies, under applied torsion in the counter-clockwise direction. The intial dislocations glide toward each other due to torsion; then there occurs a sequence of two dislocation reactions, an extrusion, and a dislocation pair-unbinding as illustrated in Figure 4 of the main text. Magnitude of applied torsion gradually increases with time.
\\

\textbf{Video 4.} A sequence of reactions in a tubular crystal under alternating sign of torsion that causes reorientation of dislocations.  

\end{document}